\documentclass[superscriptaddress,twocolumn,showpacs,preprintnumbers,prb]{revtex4}
\usepackage{graphicx}
\usepackage{times}
\usepackage{epsfig}

\begin{document}

\def\salto{\vskip 1cm}
\def\lgr{\langle\langle}
\def\rgr{\rangle\rangle}
\def\lan{\langle}
\def\ran{\rangle}

\title{ Interference Effects in the \\
Conductance of Multi-Level Quantum Dots}

\author{C.A. B\"usser, G.B. Martins, K.A. Al-Hassanieh,
Adriana Moreo, and Elbio Dagotto}
\affiliation{National High Magnetic Field Lab and Department of Physics,
Florida State University, Tallahassee, FL 32306}
\date{\today}

%\maketitle

\begin{abstract}
Using exact-diagonalization techniques supplemented by a Dyson equation
embedding procedure,
the transport properties of multilevel quantum dots are investigated in the Kondo 
regime. The conductance can be decomposed into the contributions
of each level. It is shown that these channels can carry a different
phase, and destructive interference processes are observed
when the phase difference between them is $\pm\pi$.
This effect is very different from those observed in bulk metals with
 magnetic impurities, where the phase differences play no significant role.
The effect is also different from other recent studies of interference 
processes in dots, as discussed in the text. In particular, no external magnetic
field is here introduced, and the hopping amplitudes dot-leads for all levels
are the same. However, 
conductance cancellations induced by interactions are still observed. 
Another interesting effect reported here is the formation of localized
states that do not participate in the transport.
When one of these states crosses the Fermi level, the electronic occupation
of the quantum dot changes, modifying the many-body physics of the system
and indirectly
affecting the transport properties. Novel discontinuities between two
finite conductance values can occur as the gate voltage is varied, as
discussed here.
\end{abstract}

\maketitle

%\begin{multicols}{2}  

\section{ Introduction}

Among the several interesting potential applications of
nanostructures and nanodevices is their role as
small-size test ground for the analysis of the influence of many-body
interactions in the physics of strongly correlated 
electronic systems. These correlation
effects are believed to be of crucial important in many bulk materials,
such as high-temperature superconductors and manganites, but there is
no universal agreement on what formalism is the most appropriate for their study.
Verifying the accuracy of  simple model Hamiltonians in nanoscale systems 
can provide an important
test of the overall validity of the theoretical
formalism. Moreover, several compounds are believed to spontaneously form
a complex pattern of nanoislands \cite{book} 
and, as a consequence, accurate studies of a small number
of atoms can produce interesting information even for bulk materials.

In addition, new and novel
effects can and do emerge from the quantum behavior of electrons in small
environments, which are not obviously present in bulk systems. 
If quantum dots are considered as parts of electronic circuits, it
is important to understand the new properties that may arise
due to the discrete nature of the quantized
states, the influence of interactions and spin, and 
the coherence of the wave functions in small systems that requires a fully
quantum description of the problem. It is only recently that 
this complicated physics is starting to be unveiled and novel effects
have already been found. A typical case
is provided by the Kondo effect\cite{Hewson}, which in the bulk produces an
increase in the resistance of a sample below the Kondo temperature
$T_{\rm K}$. On the contrary, in 
mesoscopic systems such as quantum dots, the same effect
actually leads to a $decrease$ in the resistance since the Kondo resonance
provides a new channel for transport \cite{Glazman,DGoldhaber}.

The analysis of dots is not a mere academic problem. 
For example, transport properties at low temperatures across quantums dots (QDs)
have received considerable attention in the last few years mainly due the
possibility to use these structures in quantum computing and spintronics.
More specifically,
two coupled quantum dots (2QD) have been
proposed to represent tunable
qubit circuits  \cite{DLoss}, where the spin interaction
between the electrons in the dots
defines the two states of the qubit. 
For all these reasons, the study of interacting electrons in nanosystems and
its associated quantum transport
are very important for progress in condensed matter physics in general.

In several previous studies it has been
usually $assumed$ that, e.g., the 2QD have only one localized spin
per dot, since only one level is assumed active in each dot. 
For the success of these proposals, 
it is important to understand the role played by other
levels in the QDs -- i.e. considering $multilevel$ dots -- providing part of the
motivation for the work presented here.
In addition, 
to properly perform these subtle studies it is crucial to carry out the calculations
using $unbiased$ techniques that do not assume properties of familiar states
that arise from previous experience with bulk materials. Concepts
such as Fermi liquids are and must be challenged in this context,
since there is no reason to expect that nanoscopic entities will
admit a similar description. Ideally, the calculations should make the less possible
number of assumptions, and in this framework it is important to
develop and use suitable $numerical$ techniques to handle these complicated
fully-quantum problems accurately. For this reason, the calculations discussed
below were carried out with a powerful numerical method that has been
recently developed.

Other interesting investigations in this context have
proposed the use of quantum dots as $interferometers$ in the Aharanov-Bohm
(AB) geometry with external magnetic fields applied \cite{RLopez,DBoese}. 
In general, previous efforts have considered 
a magnetic flux enclosed  in two possibles
paths for the electrons. In these studies, the tunneling matrix-elements 
are defined with an explicit phase factor that represents the magnetic flux.
These investigations have shown that narrow $dips$ 
in the conductance could appear as a 
consequence of phase interference in the  AB circuit, for electrons moving
through different pathways. Other investigations have also
found conductance cancellations for the case of
two quantum dots, individually connected to leads 
and among themselves \cite{Claro}.
This effect is present even using one-body interactions in the formalism,
and it is caused by interference effects between two different paths in
the geometry of the problem.

One additional 
important  motivation for the analysis discussed in the present paper is to
find alternative sources of interference among wave functions that can
also lead to dips in conductance, {\it in the absence of external fields
and also in the absence of obviously distinct paths}.
Effects of this variety were recently addressed in multilevel dots by
Kim and  Hershfield \cite{Hershfield2}, where the 
consequences of
having tunneling matrix elements with different signs in a
multilevel QD and in an AB circuit were discussed. These authors showed 
that the matrix-element signs are important to determine the global phase of
the electrons when they cross the QD. 
Also, arrays of odd-number interacting single-level
quantum dots have interesting 
zeros in the conductance as the gate voltage varies \cite{busser2}.

In the investigations presented here,
we consider a system composed of one or more quantum
dots  -- each with two active levels
and including the Hubbard many-body interactions -- coupled 
to ideal single-channel leads. In gate-voltage regimes where  there is
 only one electron in the QD, the usual
Kondo physics \cite{Glazman} can be found in the low temperature
regime. The coupling between 
the spin in the QD and the spins in the leads creates a resonance at 
the Fermi level that contributes to the transport.
When a second electron enters in the QD a more interesting physics
is observed since now the total spin of the system (singlet or triplet) 
plays an important role \cite{Sasaki,Glazman2,Guta}. 
One of the main results of our investigations is that the two levels
of the dots studied here can act as different channels that individually contribute
to the overall conductance with their own phase. This phase 
plays a role similar to that of a magnetic field in the AB interferometer,
and it can induce constructive or destructive interferences.
Below, it is explicitly shown that 
a phase diference of $\pi$ can exist between the two channels,
even if there is no explicit difference between their hopping amplitudes
and Coulombic interactions. Only a tiny difference in the energies of
the two levels is needed for the effect to develop, as well as a nonzero
Hubbard coupling which is required to produce the Kondo physics.
It should be remarked that
this phase difference is different from the phase messured
in some QD experiments \cite{Yacoby} since it is $internal$ to the QD.
A magnetic field is not necessary to observe this
interference.

Others recent investigations have also focussed on systems that
present interference processes even without magnetic
fields or explicitly different paths. For instance, Hofstetter and
Schoeller analyzed lateral quantum dots near the {\it singlet-triplet}
degeneracy point that occurs from the competition of the Hund rule's coupling,
that favors the triplet, and the nonzero energy difference between levels, that
favors a singlet state \cite{WHofstetter}. 
A dip in the conductance was observed in this region.
The study discussed in the present paper includes this singlet-triplet
transition regime as a special case, but mainly focusses on a different variety of
interference effects which are present even for a very small energy 
difference between the levels in the quantum dots. Our main picture
is based on the notion that charge transport can occur with different
phases when electrons cross different levels of a quantum dot, rather
than a singlet-triplet competition.
Alternatively, our results can also be visualized as interference 
between two Kondo states: one with $S=1$ and the other with $S=1/2$.

The organization of the paper is the following:  
In Sec.~\ref{sec:formalism} the discussion of the formalism,
mainly the model and technique, is presented.
In Sec.~\ref{sec:one-dot} the case of one dot with two
levels is studied in detail, showing the novel interferences in transport.
The analysis of the integer-spin Kondo effect, that appears here in the
interesting region of parameters, is in Sec.~\ref{sec:S1kondo}.
It is well-known that this type of states could enhance
the conductance near  singlet-triplet transitions 
\cite{Sasaki,Glazman2}.
A possible intuitive explanation of the effect 
is given in Sec.~\ref{sec:intuitive}. An analysis of other cases
where similar effects were found is in Sec.~\ref{sec:othercases},
and the conclusions are provided in Sec.~\ref{sec:conclusions}.
Other effects analyzed in the paper include the possibility of observing
{\it localized states} near the dots. 
One such localized state in the QD can cross the
Fermi level of the system without contributing to the conductance, but
affecting the transport of other levels through the Coulomb interaction.
This is shown to lead to discontinuities 
-- as opposed to cancellations -- in the conductance vs. gate voltage,
a concept that was not discussed before in the literature to our knowledge.

%%%%%%%%%%%%%%%%%%%%%%%%%%%%%%%%%%%%%%%%%%%%%%%
%%%%%%%%%   1QD con dos niveles......
%%%%%%%%%%%%%%%%%%%%%%%%%%%%%%%%%%%%%%%%%%%%%%%
\section{One Dot with two levels: Formalism}
\label{sec:formalism}

\begin{figure}
\vskip 0.25cm
\epsfxsize=6.5cm \centerline{\epsfbox{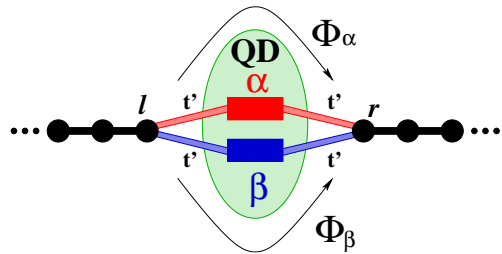}}
\vskip 0.25cm
\caption{ Schematic representation of the main system studied in this paper,
consisting of one quantum-dot with two levels $\alpha$ and $\beta$. 
The two possible paths 
carrying different phases $\Phi_\alpha$ and $\Phi_\beta$
are shown.}
\label{esquema-1QD2l}
\end{figure}

The main system that will be analyzed in this paper, and especially in this 
section, consists 
of one quantum dot with
several interacting levels, connected to ideal leads. 
It will be argued that -- mainly due to the many-body interaction --
in a transport process these levels
can carry a different phase,
as shown schematically in Fig. \ref{esquema-1QD2l},
giving rise to a complex conductance pattern including
destructive interferences (i.e. an ``internal''
Bohm-Aharonov effect appears to be dynamically generated).
To model this system, the impurity Anderson Hamiltonian 
with two levels ($\alpha$ and $\beta$) is used. The portion corresponding to 
the isolated dot is
\begin{eqnarray}
H_{\rm dot} &=& \sum_{\lambda=\alpha,\beta ~\sigma}  
U/2 ~n_{\lambda \sigma} ~n_{\lambda \bar{\sigma}}  ~+~
U'/2 ~\sum_{\sigma \sigma'}~n_{\alpha \sigma} ~n_{\beta \sigma'} ~+~ \nonumber \\
 && -J~ \sum_{\sigma \sigma'} ~c^+_{\alpha \sigma} ~c_{\alpha \sigma'} ~c^+_{\beta \sigma'}
 ~c_{\beta \sigma} ~+~ \nonumber \\ &&
  \sum_{\sigma} \left[ V_g ~n_{\alpha \sigma} ~+~
   (V_g + \Delta V) ~n_{\beta \sigma} \right],
\end{eqnarray}
where the first term represents the usual Coulomb repulsion between
electrons in the same level, which for simplicity is considered equal 
for both. The
second term represents the Coulomb repulsion between electrons in
different levels (the $U'$ notation is borrowed from standard many-orbital
studies in atomic physics problems). The third term represents the Hund coupling 
($J>0$) that favors the alignment of spins,
and the last term is simply the energy of the states 
regulated by the gate voltage $V_g$. Note
that the level $\alpha$ is assumed to be separated from $\beta$ by $\Delta V$,
and by modifying this parameter an interpolation between one- and two-level
physics can be obtained.
The dot is connected to the leads (represented by semi-infinite chains)
by a hopping term with amplitude $t'$, while $t$ 
is the hopping amplitude in the leads. More specifically,
\begin{eqnarray}
H_{\rm leads} &=& t ~\sum_{i \sigma} \left[ c_{l i\sigma}^+ ~c_{l i+1\sigma}
 ~+~  c_{r i\sigma}^+ ~c_{r i+1\sigma} ~+~\mbox{h.c.} \right], \\
H_{\rm int} &=& t' \sum_{\sigma, \lambda=\alpha,\beta} \left[ 
 c_{\lambda \sigma}^+ \left(c_{l 0\sigma} ~+~ c_{r 0\sigma} \right)
 ~+~ \mbox{h.c.} \right] ,
\end{eqnarray}
where $c_{l i\sigma}^+$ ($c_{r i\sigma}^+$ ) creates electrons at
site $i$ with spin $\sigma$ in the left (right) contact. Site ``0'' is
the first site at the left and right of the dot, for each half-chain.
The total Hamiltonian is
\begin{equation}
H_{\rm T} = H_{\rm dot} ~+~ H_{\rm leads} ~+~ H_{\rm int}.
\end{equation}
Note that for $V_g$=$-U/2 -U' +J/2 - \Delta V/2$, the Hamiltonian
is particle-hole symmetric.

Using the Keldysh formalism \cite{Meir-cnd}, 
the conductance through this system can be written as
\begin{equation}
\sigma =  \frac{e^2}{h} ~|t^2~G_{lr}(E_{\rm F})|^2  \left[ \rho(E_{\rm F}) \right]^2,
\label{conductancia}
\end{equation}
where $G_{lr}(E_{\rm F})$ is the Green function that
corresponds to moving an electron from the left
lead to the right one, $E_{\rm F}$ is the
Fermi energy, and $\rho(E_{\rm F})$ is the density of states of
the leads (assumed the same left and right, for simplicity):
\begin{equation}
\rho(\omega) = \frac{-1}{\pi} \mbox{Im} \left\{ 
\frac{(\omega+i\eta) - \sqrt{(\omega+i\eta)^2 - 4t^2}}{2t^2}
\right\}.
\end{equation}

%%%%%% Calculo de la fase:
To use Eq.\ref{conductancia}, $ |G_{lr}|^2$  must be calculated.  
This quantity has
two contributions, one corresponding to transport through path 
"$\alpha$" and the
other through path "$\beta$". From the equations of motion \cite{EOM},
which correspond to an exact expansion of the Green function \cite{noteEOM}, 
it can be shown that
\begin{eqnarray}
G_{lr} &=& \tilde{g}_l ~t' ~G_{\alpha r} + \tilde{g}_l ~t' ~G_{\beta r} ,
 \label{Glr} 
%% \\  &=& M_\alpha e^{i\Phi_{\alpha}} + M_\beta e^{i\Phi_{\beta}}
   \label{fases}    
\end{eqnarray}
where $\tilde{g}_l$ are the Green function 
of the first left-contact and
$G_{\alpha r}$ ($G_{\beta r}$) is the  dressed Green functions 
to move from the state "$\alpha$"  ("$\beta$") to the right-contact.

The two terms in Eq.\ref{fases}, involving
$G_{\alpha r}$ and $G_{\beta r}$, can be evaluated 
independently.  To calculate the phase difference 
the following expression was used 
\begin{equation}
i~\Delta\Phi = \log \left\{ \frac{G_{\alpha r}}{G_{\beta r}} 
~~\frac{|G_{\beta r}|}{|G_{\alpha r}|} \right\} .
\label{deltaphi}
\end{equation}

The conductance is proportional to
$|G_{lr}|^2 $ and, using Eqs. \ref{conductancia}, \ref{Glr} and 
\ref{deltaphi}, it can be written as 
\begin{equation}
\sigma = \sigma_\alpha ~+~ \sigma_\beta ~+~ 
\sqrt{\sigma_\alpha~\sigma_\beta} ~\cos{\Delta\Phi } ,
\label{cond-2path}
\end{equation}
where $\sigma_\alpha$ and $\sigma_\beta$ are the partial 
conductances given by
\begin{eqnarray}
\sigma_\alpha &=& \frac{e^2}{h} \left[t^2\tilde{g_l}~t'~\rho(E_{\rm F})\right]^2
 ~~|G_{\alpha r}(E_{\rm F})|^2 , \\
\sigma_\beta  &=& \frac{e^2}{h} \left[t^2\tilde{g_l}~t'~\rho(E_{\rm F})\right]^2
 ~~|G_{\beta r}(E_{\rm F})|^2  . 
%%  \\
%%\tilde{\sigma}&=& \frac{et}{h} \left[\tilde{g_l}~t'~\rho(E_{\rm F})\right]^2
\end{eqnarray}

An exact cancellation of the conductance can occur when 
$\sigma_\alpha$=$\sigma_\beta$
and $\Delta\Phi$=$\pm \pi$ as a consequence of a destructive interference 
process. In the same way, the situation with $\Delta\Phi$=$0$ and
$\sigma_\alpha$=$\sigma_\beta$ can be considered as a constructive 
interference.

The zero-temperature Green functions for this problem were calculated 
using an exact-diagonalization (Lanczos) method\cite{Elbio}
to obtain the ground
state and the Green functions of a small cluster containing the
many-body interactions. This is followed by an ``embedding'' procedure to reestablish
the rest of the leads \cite{meth1,busser}.
The first step in this technique is to separate the system in two portions. One
of them is a cluster containing the levels $\alpha$ and $\beta$ (interacting
region) and also including
the first few sites left and right for each contact.
The total number of levels in the cluster is denoted by $L$. 
For one two-level quantum dot, this corresponds to $L-1$ sites,
since one of them has 2 active levels. 
Since Kondo physics
will be important in this analysis, treating exactly not only the dot but
also some sites belonging to the leads allows for a proper quantitative 
consideration of the spin-singlet Kondo ``cloud''.
The other portion in the problem corresponds to the rest of
the contacts. Denoting by $\hat{g}$ the exactly calculated 
Green functions (at $T$=$0$) of the cluster, 
the rest of the system can be incorporated using the Dyson equation 
$\hat{G}$=$\hat{g} + \hat{g}~\hat{t}~\hat{G}$, where $\hat{t}$ are the
matrix elements that connect the cluster with the contacts,
and $\hat{G}$ is the dressed Green function.

To take into account possible charge fluctuations inside the cluster,
a linear combination of $\hat{g}$'s calculated with different
number of particles is considered. For this purpose,  
$\hat{g}^p$=$(1-p) \hat{g}_n + p~\hat{g}_{n+1}$ is defined, 
where $\hat{g}_m$ is the
Green function for $m$ particles. With the dressed Green function
$\hat{G}^p$, the total charge inside the cluster after the embedding 
is calculated using
$Q^p$=$-1/\pi \int_{-\infty}^{E_{\rm F}} \sum_i \mbox{Im}~G_{ii}^p(\omega) d\omega$.
On the other hand, the charge in the cluster before the embedding process
was $q^p$=$(1-p) n + p (n+1)$. These two quantities must be equal 
($Q^p = q^p$), defining a self-consistent equation for $p$.  

One of the important scales in this problem is the broadening of 
levels of the dot due to the interaction with the leads. This
energy scale, $\Gamma$=$\pi t'^2 \rho(E_{\rm F})$\cite{Hewson},
defines the Kondo temperature for the case of a spin-$1/2$
inside the quantum dot \cite{Hewson}.
The coupling values used in this paper are (unless otherwise stated) 
$U$=$0.5t$, $U'$=$2/3U$, $J$=$U-U'$, and $t'=0.2t$,
that allows for a fast convergence with the cluster size \cite{busser}.
Fixing the values for the Coulombic interaction couplings
is only for simplicity,
to avoid exploring a vast parameter space. Considering $U'$ to be
smaller than $U$ is reasonable since the latter is expected to be
the largest Coulombic interaction. The equation for $J$ is borrowed
from atomic physics problems\cite{manganites-review}, 
and it gives a reasonably small value for the
Hund interaction in units of $U$. In addition,
we selected the ratio $\Gamma/U$ to be close to the experimental
values\cite{Golhaber2}.

\section{Conductance Cancellations in Multilevel Dots}
\label{sec:one-dot}

\begin{figure}
\epsfxsize=7.5cm \centerline{\epsfbox{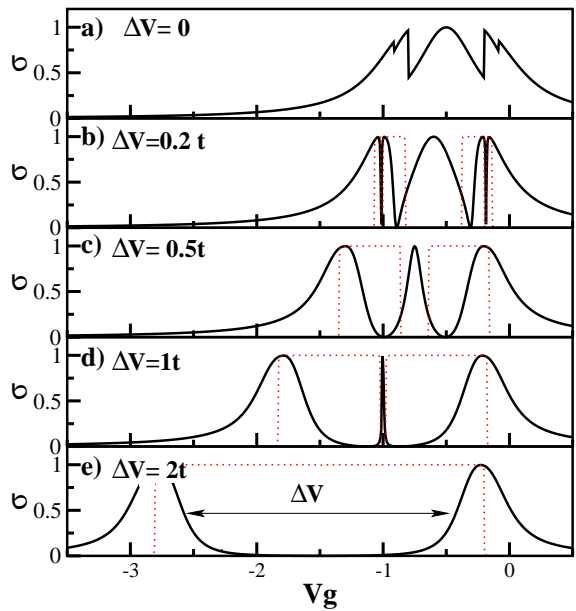}}
\caption{ Conductance for five different values of $\Delta V$.
The conductance (in units of $e^2/h$) is shown in solid lines and the phase difference 
(divided by $\pi$) in dotted lines.
When $\Delta V$ is the largest energy ($\Delta V$=$2t$ (e))
two structures similar to those found in a single-level QD,  separated
by $\Delta V$, appear in the conductance. The case (d) is similar,  
but since $\Delta V$ is not large enough, an extra thin peak appears
around $V_g=-1$ (this peak is shown with more detail in Fig. \ref{detail1.0}).
When $\Delta V$ decreases,
these structures began to interact giving rise to 
dips in the conductance, as shown in (b) and (c)  for the cases 
$\Delta V$=$0.2t$ and $\Delta V$=$0.5 t$
($\Delta V$=$0.2t$ is studied in more detail in Fig. \ref{DV0.2}). 
Finally, in the limit $\Delta V$=$0$,
the conductance dips are transformed into discontinuities. }
\label{fg1-1qd2l}
\end{figure}

Using the Lanczos method to evaluate the Green functions of a
small cluster followed by the embedding process described 
in the previous section, we have calculated the conductance, 
the charge of the QD, and the phase difference Eq.\ref{deltaphi}
as a function of $V_g$. 
Unless otherwise stated, we used the smallest possible cluster size
$L$. For the case analyzed in this section this smallest size is $L=4$
that correspond to two ``lead'' sites and two levels in the dot.
In the last part of this section we analyze the convergence of
the results increasing $L$.
In Fig. \ref{fg1-1qd2l}, we show the conductance for
several values of the energy separation $\Delta V$ between the two 
levels of the dot. 
At large $\Delta V$ (Fig. \ref{fg1-1qd2l}(e))
the levels are well separated and the interaction between
them is weak. The conductance shows just two wide peaks associated
with the standard Kondo effect for each level. Note that the width
of these peaks is $U$, and with increasing temperature
this broad peak will transform into two sharper peaks corresponding
to the Coulomb blockade. This physics has been discussed at length
in previous publications \cite{busser,Guta,busser2}, and it basically amounts to the behavior
of single-level QDs, which is natural at large $\Delta V$.

Far more interesting results are obtained by
reducing $\Delta V$. For instance, in Fig. \ref{fg1-1qd2l}(d)
a thin central-peak is already observed.
This peak seems to be the consequence of a constructive interference
between the two possible channels, in a region 
where a state with total spin $S_{\rm D}$=$1$ is formed, as will be discussed
more extensively below.
Upon further decreasing $\Delta V$ (Figs. \ref{fg1-1qd2l}(b) and (c) ) 
the two largest structures 
began to interact with the thin central peak, which increases its width,
and a more complex pattern emerges.
With the levels of the dot getting closer it is more likely to
have them populated at the same time, and
due to the Hund coupling the state with $S_{\rm D}$=$1$ has a high probability
for the case of two electrons in the dot,
giving rise to a spin-1 Kondo effect within the central peak.
The other two peaks are associated 
with $S_{\rm D}$=$1/2$ 
and 1 and 3 electrons in the dot. For $\Delta V$$\leq$0.5$t$,
dips in the conductance are clearly observed, and later we will show that
these cancellations are related to interference processes.
Finally, at $\Delta V$=$0$ (Fig. \ref{fg1-1qd2l}(a)) 
several {\it discontinuities} appear in the conductance.
It will be argued that these discontinuities are a consequence of
the existence of $localized$ states at the dot. 
These states are not of direct relevance for the transport of charge, but
when the gate voltage moves them below the Fermi level of the leads
they simply increase the charge of the quantum dot, which indirectly
affects transport through the Coulombic interaction.
Note that, for $\Delta V$=$0$,  the phase difference between 
levels $\alpha$ and $\beta$ is zero, since the Green functions 
$G_{\alpha 2}$ and $G_{\beta 2}$ are equal. Thus, no destructive interferences
are observed, but discontinuities can still occur.

\begin{figure}
\epsfxsize=8cm \centerline{\epsfbox{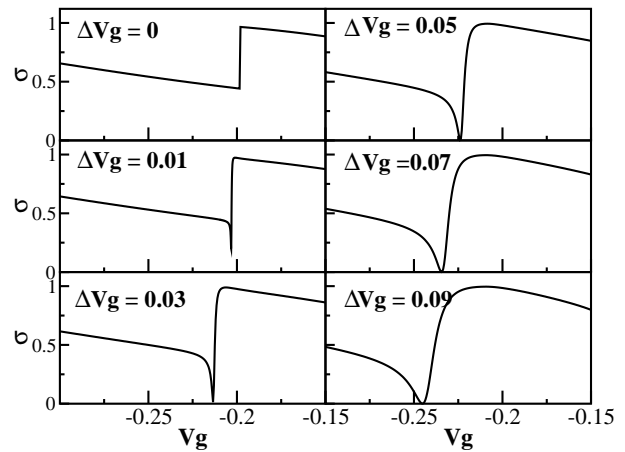}}
\caption{ Conductance (in units of $e^2/h$) in a narrow range 
of $V_g$ and $\Delta V$. The discontinuity 
at $\Delta V$=$0$ is transformed into a dip for finite values
of $\Delta V$. }
\label{varios}
\end{figure}

\begin{figure}
\epsfxsize=8cm \centerline{\epsfbox{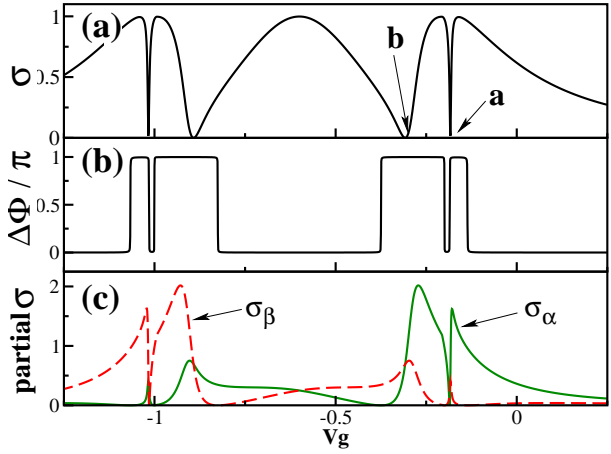}}
\caption{ Conductance (in units of $e^2/h$) (a), 
phase difference (divided by $\pi$) (b),
and partial conductances (c) for $\Delta V$=$0.2 t$. 
The dips, indicated by ``a'' and ``b'' in (a),
are in regions where the phase difference jumps to $\pi$ 
producing an exact cancellation when 
$\sigma_\alpha$=$\sigma_\beta$.}
\label{DV0.2}
\end{figure}

\begin{figure}
\epsfxsize=8cm \centerline{\epsfbox{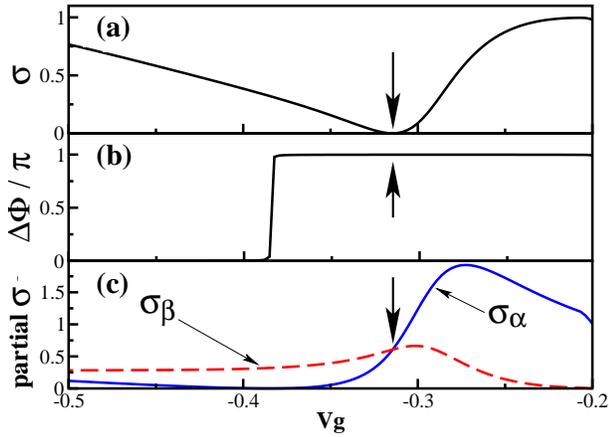}}
\caption{ Conductance (in units of $e^2/h$), phase difference, and partial 
 conductances for $\Delta V$=$0.2 t$ for the ``b'' dip of Fig. \ref{DV0.2}(a). 
 Thicks arrows shows the gate potential where the conductance cancells, 
 $\Delta \Phi = \pi$ and $\sigma_\alpha = \sigma_\beta$.}
\label{DV-dip0.2}
\end{figure}

\begin{figure}
\epsfxsize=8cm \centerline{\epsfbox{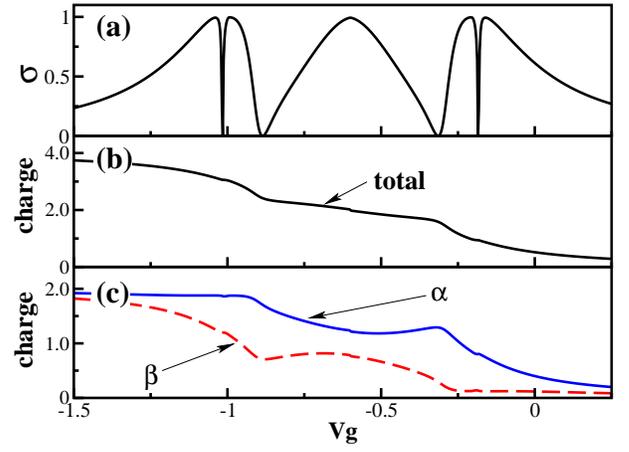}}
\caption{ Conductance (in units of $e^2/h$), total charge of the QD, 
and charge at each
level for the case $\Delta V$=$0.2t$. The charge of the level $\alpha$ 
is indicated with a continuos lines, while the charge of $\beta$
is shown with a dashed one. The central peak emerges in the 
regime where there are two electrons in the dot, in different levels.}
\label{DV-carga0.2}
\end{figure}

In Fig. \ref{varios}, it is shown how one of the conductance discontinuities 
is rapidly transformed into a dip as $\Delta V$
increases from zero. In Figs. \ref{DV0.2}, \ref{DV-dip0.2} and \ref{DV-carga0.2}
we can observe in detail the case with $\Delta V$=$0.2t$.
Figure \ref{DV0.2} shows the conductance together
with the phase difference and the partial conductance.
Clearly, the zeros of the conductance are
in regions where $\Delta \Phi$=$\pi$. Then, 
from Eq.\ref{cond-2path}, the dips
must occur when the partial conductance $\sigma_\alpha$
and $\sigma_\beta$ take the same values.
More detail for dip "b" is
shown in Fig. \ref{DV-dip0.2} 
where thick arrows point to the zero in the conductance and
the location where $\sigma_\alpha = \sigma_\beta$.
The dot charge for the same $\Delta V$ is shown in Fig. \ref{DV-carga0.2}.
The central peak in the conductance occurs in the regime where
two electrons are in the dot, in different levels.
It can also be observed  that the total charge at the QD ($\sum_\sigma~n_{\alpha \sigma} 
+ n_{\beta \sigma}$) does not show the characteristic sharp plateauxs of
the Coulomb blockade regime: here the electrons are entering in the QD
in a smoother way since the Coulomb interaction is comparable 
to the hopping amplitude $t$ (Fig. \ref{DV-carga0.2}(b) ). 

\begin{figure}
\epsfxsize=8cm \centerline{\epsfbox{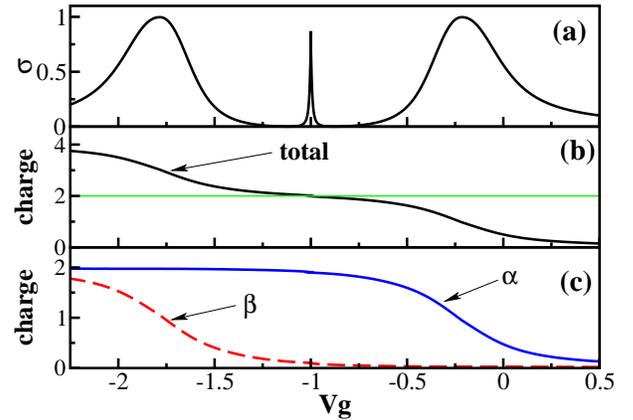}}
\caption{ Conductance (in units of $e^2/h$) and charges for the case $\Delta V$=$t$
as a function of the gate potential. These results should be compared against those
of Fig. \ref{DV-carga0.2}}
\label{dV=1}
\end{figure}

Similar results but for the case of $\Delta V$=$t$ are shown in Fig. \ref{dV=1}.
In this situation, the large $\Delta V$ well
separates the levels $\alpha$ and $\beta$, and as a consequence one of the
levels is charged almost completely before some charge
starts to enter into the other level (Fig. \ref{dV=1}(c)). The latter occurs
at the $V_g$ where the conductance peak is observed.

\begin{figure}
\epsfxsize=8cm \centerline{\epsfbox{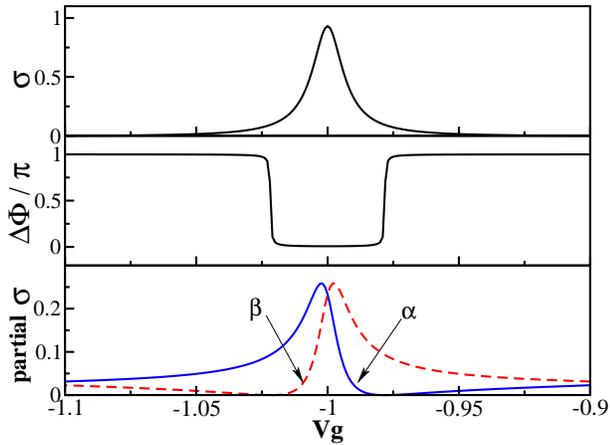}}
\caption{ Conductance (in units of $e^2/h$), phase difference and partial 
conductances for $\Delta V$=$t$ for the central peak shown in
Fig. \ref{fg1-1qd2l}(d).
The partial contributions to the conductance ($\sigma_\alpha$ 
and $\sigma_\beta$) have the same shape as in Fig. 
\ref{DV0.2} but now the phase difference between them is zero. 
This phase difference creates a peak where in the other situation 
appears a dip.}
\label{detail1.0}
\end{figure}

To better understand the physics associated with the thin 
central peak for the case $\Delta V$=$t$ shown in Fig. \ref{fg1-1qd2l},
the conductance and the phase for this case is presented in more detail in
a narrow $V_g$ range in Fig. \ref{detail1.0}.
We can observe that the partial conductances $\sigma_\alpha$
and $\sigma_\beta$ are individually small ($< 0.25$) but as the phase
between the two channels is 0 they contribute in a constructive way,
inducing a peak in the total conductance.

\begin{figure}
\epsfxsize=8cm \centerline{\epsfbox{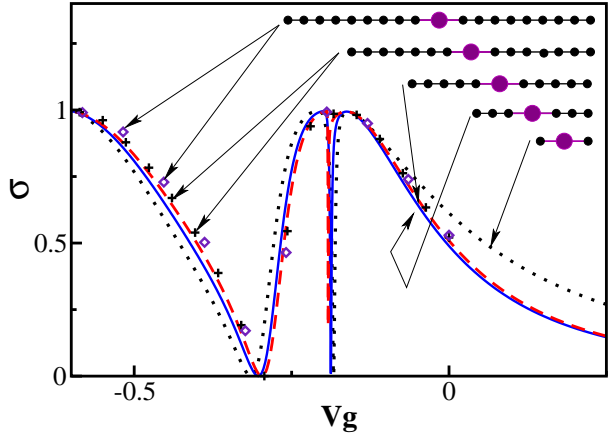}}
\caption{ Convergence of the conductance with the size of 
the exactly-solved cluster $L$ for
the case $\Delta V$=$0.2t$.  Dotted lines are for
$L=4$, continuous thin lines for $L=8$, and dashed lines for
$L=12$. Extra points for $L=16$ and $L=20$ are shown 
with crosses and diamonds, respectively.}
\label{cnvg-extra5}
\end{figure}

Our results have been obtained solving exactly a relatively small cluster 
and, then, approximately reaching the bulk limit using the Dyson equation.
It is important to verify that our results are robust if the cluster 
size increases.
In Fig. \ref{cnvg-extra5}, the convergence 
of the conductance with increasing cluster size
at $\Delta V$=$0.2t$ is shown. Remarkably,
the diference between $L$=$8$ and
$L$=$20$ is less than 0.5\%, showing that the results presented here
are nearly exact. 
The fast convergence is a consequence of the local characteristics of
the dot, namely all the important physics effects are contained 
in the small cluster solved exactly.
The fast size convergence shows
that even using the smallest cluster, the results 
are already qualitative correct.

\begin{figure}
\epsfxsize=8cm \centerline{\epsfbox{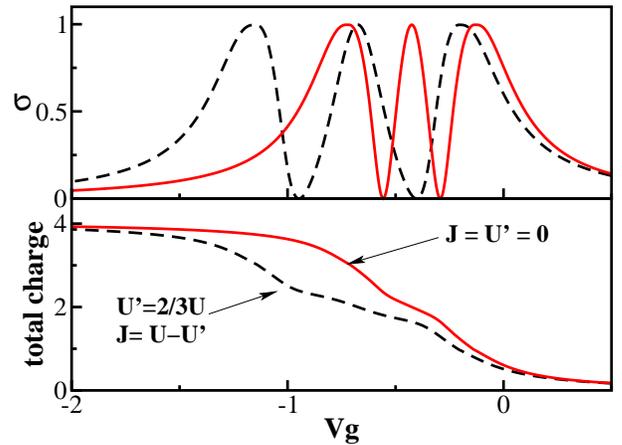}}
\caption{ Study of the conductance and total charge when $U'$ and
$J$ are not active. Shown are results for 
the case $U'$=$J$=$0$
(solid line) compared with the previously shown results at $U'$=$2/3U$
and $J$=$U-U'$, both at  $\Delta V$=$0.35$.
The effects are qualitative the same in both cases. }
\label{JUp=0}
\end{figure}

\begin{figure}
\epsfxsize=8cm \centerline{\epsfbox{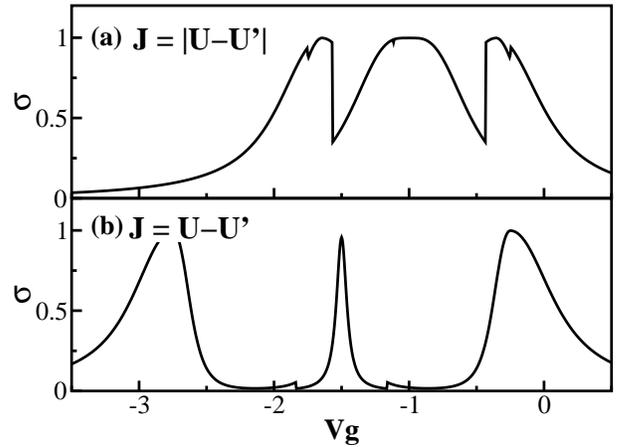}}
\caption{ Conductance (in units of $e^2/h$) with large $U'$=$2t$.
In (a), it is shown the case with a ferromagnetic 
Hund coupling $J$=$|U-U'|$ and in (b) 
with antiferromagnetic $J$=$U-U'$. Both figures
are for $\Delta V$=$0$.}
\label{largeUp}
\end{figure}

It is important to remark that the cancellations observed here
do not crucially depend on the couplings $U'$ and $J$, since they
are present in our results even for $U'$=$J$=0, as shown in 
Fig. \ref{JUp=0}. Also for completeness,
in Fig. \ref{largeUp} the case of a large
$U'$=$2t$ is shown. Even though the situation $U'$$>$$U$ is unphysical
-- since the Coulomb repulsion 
in the same level should be larger than for electrons
in different levels -- it is interesting to note that the
$U'$ energy plays a role similar to that of $\Delta V$.
Since the energy to have a second electron in 
$\beta$ is related to $U'-J+\Delta V$, for a large $U'$ 
(with a $J$ with the same sign as considered before)
the system reacts as with large $\Delta V$.
Figure \ref{largeUp} contains two differents cases for
$\Delta V$=$0$, one of them with
a ferromagnetic Hund coupling $J$$=|U-U'|$$>$$0$ (a) and the
other with an antiferromagnetic coupling $J$=$U-U'$$<$$0$ (b).
In the first case, since the Hund coupling is proportional to $U'$,
we observe the three-peaks structures for
the conductance with a central-peak associated
to the $S_{\rm D}$=$1$ states.
In the second case, $U'$ and $J$ try to eliminate the $S_{\rm D}$=$1$
states, and the conductance has a similar structure as shown in
the previously study example with $\Delta V$=$t$ 
(see Fig. \ref{fg1-1qd2l}(d)).

\section{ Integer-Spin Kondo Effect}
\label{sec:S1kondo}

Recently, transport through a dot connected to two leads
has been experimentally studied as a function of the dot level separation
($\Delta V$), tuned by an external magnetic field. For a dot 
with an even number of electrons, it has been shown that 
the Kondo correlations formed between the $S_{\rm D}$=$1$ triplet state of
the dot and the spins of the leads has a strong influence on the 
current \cite{Sasaki}.
This problem was theoretically studied using a mapping onto a
two-impurity Kondo model \cite{Glazman2,Guta}. 

\begin{figure}
\epsfxsize=8cm \centerline{\epsfbox{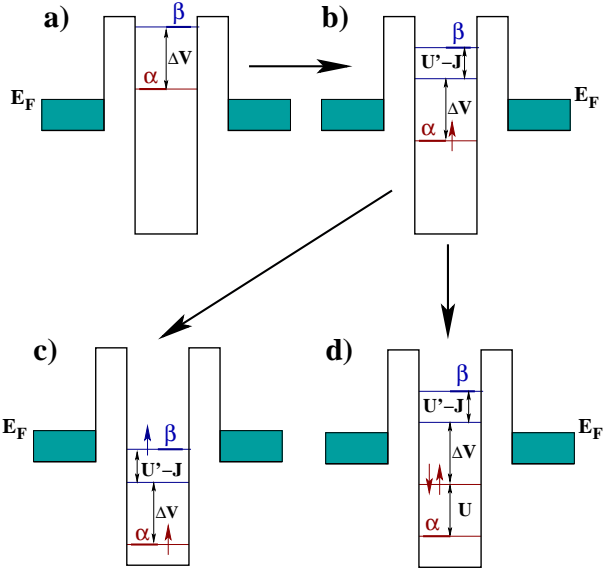}}
\caption{ Schematic representation 
of the different situations that could occur
in a  two-level QD as the gate voltage changes. For more details see
the text.}
\label{esq-spin1}
\end{figure}

Figure \ref{esq-spin1} shows schematically the different situations that
could arise as  the number of electrons inside the QD is changed using
the gate voltage. Starting from
the upper left (a), we have the case where the gate potential $V_g$ is
above the Fermi level and there are no electrons inside the QD.
Level $\alpha$ is separated 
from level $\beta$ by the energy $\Delta V$. Decreasing 
$V_g$, as shown in Fig.\ref{esq-spin1}(b), one electron populates
level $\alpha$, and the energy to add an extra electron in
$\beta$ will be $\Delta V + U' - J$, since an
extra energy $U'$ is paid due to the Coulomb repulsion and a $-J$ energy
due to the spin coupling. The system now
is in the Kondo regime for transport if $T$$<$$T_{\rm K}$. Decreasing 
even further $V_g$, a second electron could enter the dot.
The spin of this second electron defines the transport properties
since the total spin of the QD  can be $S_{\rm D}$=$1$ (with a Kondo resonance
for temperatures below $T_{\rm K(S=1)}$) or $S_{\rm D}$=$0$ with no Kondo
effect. In the right-lower frame (d) 
the second electron entered in the lower energy state $\alpha$.
The energy of this configuration will be of the form 
$E_{S=0}$=$2V_g + U$. On the other hand, if
the second electron enters  to the dot with the
same spin as the first one forming an $S_{\rm D}$=$1$ state, we will have
the case showed in the left-lower frame (c). The energy
of this state in the Kondo regime will be of the form
$E_{S=1}$=$2V_g + \Delta V + U' - J - T_{\rm K(S=1)}$ \cite{Guta}. 
For $\Delta V$$>$$U-U'+T_{\rm K(S=1)}$ we expect a transition 
from the $S=1$ state (right-lower frame) to the $S=0$ 
(left-lower frame).

Note that the spin of the dot $S_{\rm D}$ is not a 
well-defined quantum number, since the QD is not isolated. 
In the $L=4$ cluster, we found that the ground
state used for the calculations has a spin $S_{\rm D}$=$0.7426$ 
(calculated from the relation 
$\lan \vec{S_{\rm D}}^2 \ran_{\rm cluster}$=$S_{\rm D}(S_{\rm D}+1)$), at
$\Delta V$=$0.2$ and $V_g$=$-0.6$ (particle hole symmetry). 
Since the number of particles at this gate potential is 2,
we deduce that the ground state can be expressed as a combination
of states with $S_{\rm D}$=1 and 0, with a weight of 86 percent
for $S_{\rm D}$=$1$. In the case $\Delta V$=$t$, we obtain for
the particle-hole symmetric gate potential a total spin $S_{\rm D}$=$0.352$,
which corresponds to a 60 percent chance of having $S_{\rm D}$=$1$.

\begin{figure}
\epsfxsize=8cm \centerline{\epsfbox{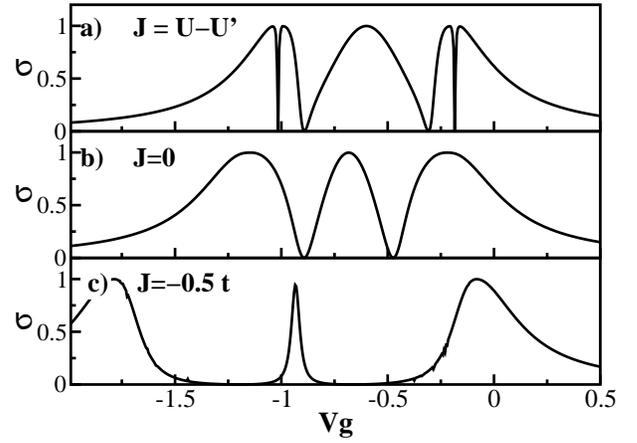}}
\caption{ Conductance (in units of $e^2/h$) for differents 
$J$ couplings.
(a) corresponds to the case $J$=$U-U$,
(b) is at $J$=$0$, and (c) is for a large negative case $J$=$-0.5t$.
The central peak disappears when the
ferromagnetic spin correlations are eliminated at even larger
negative $J$. More detail can be found in the text.}
\label{differents-J}
\end{figure}

One way to confirm that the central peak in Fig. \ref{DV0.2}(a)
indeed originates in a $S_{\rm D}$=$1$ state is by reducing the value of 
$J$, making less probable the ferromagnetic coupling between
levels $\alpha$ and $\beta$. Figure \ref{differents-J} shows the conductance
for the case shown in Fig. \ref{DV0.2}(a) plus two differents
cases with the same values of $U$, $U'$ and $t'$ but different
$J$'s. In (b) with $J$=0 we observe a narrow central peak
showing that the effect persists due to the
$T_{\rm K(S=1)}$ energy. For negative values of $J$, as shown in (c),
the ferromagnetic state is even less probable and the central 
peak is more narrow. This peak must dissapear for larger negatives
values of $J$ (i.e. $|J| >> T_{\rm K(S=1)}$).   
All this evidence suggests that it is the spin-one Kondo effect
that is responsible for some of the effects discussed here.

%%%%%%%%%%%%%%%%%%%%%%%%%%%%%%%%%%%%%%%%%%%%%%%
%%%% EXPLANATION OF THE EFFECT
%%%%%%%%%%%%%%%%%%%%%%%%%%%%%%%%%%%%%%%%%%%%%%%
\section{ Understanding the interference in the conductance}
\label{sec:intuitive}

The previous analysis showed that the origin of the 
three peaks in the conductance can be traced back to a Kondo 
effect that occurs for 1, 2, and 3 electrons inside the dot.
However, it remains to be understood 
what causes the conductance cancellation between them.
It is expected that the two paths -- through $\alpha$ and $\beta$ --
are responsible for the interference, 
but a more detailed explanation would be desirable.

\subsection{Origin of the $\pi$ Phase Difference}

\begin{figure}
\epsfxsize=8.cm \centerline{\epsfbox{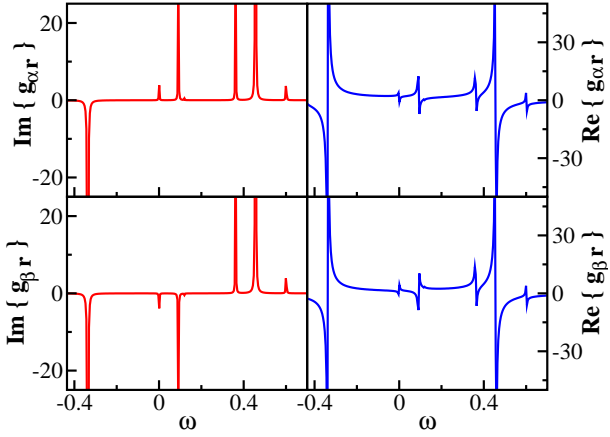}}
\caption{ Density-of-states of the bare $L=4$ cluster, for the
cases $\Delta V$=$0.05t$ and $V_g$=$-1.082t$. 
The Fermi level lies at $\omega$=$0.0$.
Note the relative change of the weight {\it signs} for the poles near
$\omega=0$, when comparing the Green functions $g_{\alpha N}$ and $g_{\beta N}$. }
\label{cluster-fase}
\end{figure}

To understand how the phase difference arises let us first 
study the $cluster$ Green function. In Fig. \ref{cluster-fase},
the real and the imaginary parts of $g_{\alpha r}$ 
and $g_{\beta r}$ are shown
for the case of $\Delta V$=$0.05t$ and a gate potential
where the relative phase is $\pi$ ($V_g$=$0.108t $). 
Since the cluster is by definition a finite system, 
these two Green functions are 
simply a collection of a large number of poles. For $\Delta V$=$0$, 
level $\alpha$ is equal 
to $\beta$  in energy, and both Green functions must have poles at exactly 
the same positions.
We expect that for a small $\Delta V$ their features must still 
remain similar. The most important information arising 
from Fig. \ref{cluster-fase}
is that the poles at approximately the same energy corresponding to $g_{\alpha r}$ 
and $g_{\beta r}$ near the Fermi level ($\omega=0$) have $different$ $signs$ in their
weights. 
This means that those single poles have a phase difference of $\pi$, which
causes the interference.

\begin{figure}
\epsfxsize=8.cm \centerline{\epsfbox{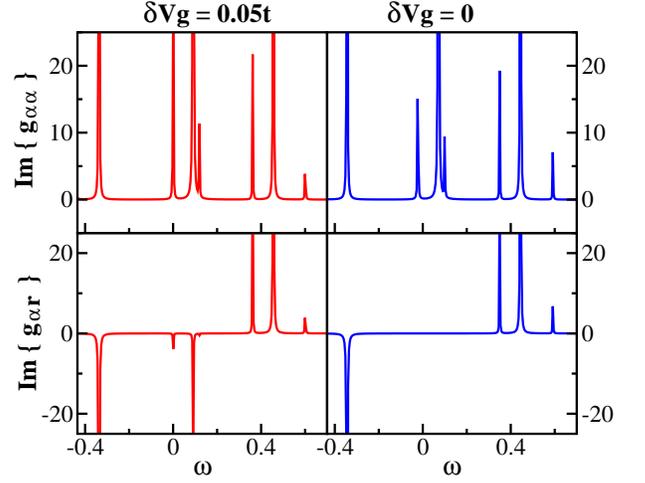}}
\caption{ Comparison between the imaginary parts of
$g_{\alpha \alpha}$ and $g_{\alpha N}$, for the cases
$\Delta V$=$0$ and $\Delta V$=$0.05t$, working at $V_g=-1.082t$ for the
$L=4$ cluster. 
The pole in $g_{\alpha N}$ near the Fermi level disappears (i.e. carries
no weight) at $\Delta V$=$0$.}
\label{cluster-dif}
\end{figure}

Other special property of the bare (cluster) Green function is shown
in Fig. \ref{cluster-dif}. In this figure, the  
imaginary parts of $g_{\alpha \alpha}$ and $g_{\alpha r}$ 
for $\Delta V$=$0.05t$ and $\Delta V$=$0 $ are shown. While
for $\Delta V$=$0.05t$ both Green functions have poles at
the same positions, at $\Delta V$=$0$ the weight of the poles
near the Fermi level for $g_{\alpha r}$ are $zero$. This is
a characteristic of a $localized$ state.

To understand these two properties we have to analyze the non-diagonal
Green functions $g_{\alpha r}$ and $g_{\beta r}$ in their Lehmann
representation \cite{JNegele}:
\begin{eqnarray}
g_{\alpha r}(\omega) &=& \sum_l \frac{\lan f| c_{\alpha \sigma} |\phi_l \ran
 ~\lan \phi_l| c^+_{r \sigma}  |f \ran}{ \omega ~-~ (E_{\rm F} - E_l) ~+~ i\eta}
 \nonumber \\
 & & + \sum_l \frac{\lan f| c^+_{r \sigma} |\phi_l \ran
 ~\lan \phi_l| c_{\alpha \sigma}  |f \ran}{ \omega ~+~ (E_{\rm F} - E_l) ~+~ i\eta} ,
\label{lehmann-a}
\end{eqnarray}
\begin{eqnarray}
g_{\beta r}(\omega) &=& \sum_l \frac{\lan f| c_{\beta \sigma} |\phi_l \ran
 ~\lan \phi_l| c^+_{r \sigma}  |f \ran}{ \omega ~-~ (E_{\rm F} - E_l) ~+~ i\eta}
 \nonumber \\
 & & + \sum_l \frac{\lan f| c^+_{r \sigma} |\phi_l \ran
 ~\lan \phi_l| c_{\beta \sigma}  |f \ran}{ \omega ~+~ (E_{\rm F} - E_l) ~+~ i\eta} ,
\label{lehmann-b} 
\end{eqnarray}
where $|f\ran$ and $E_{\rm F}$ are the ground state and its eigenenergy, 
$\{ |\phi_l\ran \}$
is a complete basis of Hamiltonian eigenstates with eigenenergies $E_l$, and
$\eta$ is an small number (typically $~10^{-7}$).

Both Green functions have poles in the same positions $\omega = \pm (E_{\rm F}-E_l)$.
For simplicity, only the first term in Eqs. \ref{lehmann-a}
and \ref{lehmann-b} will be analyzed, 
but the study presented below is qualitatively the same for the second term.
As the matrix element $\lan \phi_l| c^+_{r \sigma}  |f \ran$ is the same for
both Green functions, it is clear that they only can have a phase difference in
$\pi$ if the matrix element $\lan f| c_{\alpha \sigma} |\phi_l \ran$ has
a different sign than $\lan f| c_{\beta \sigma} |\phi_l \ran$.

This change in the sign of the matrix elements can be understood based on
the ``reflection'' symmetry between levels $\alpha$ and $\beta$. For the case
$\Delta V$=$0$, 
the transformation $\alpha \to \beta$ must leave the Hamiltonian
invariant. The operator $\hat{O}_r$ associated with this operation
commutes with the total Hamiltonian and
has eigenvalues $1$ and $-1$ (if the states are even or odd under this 
transformation). The complete basis $\{ \phi_l \}$ used in Eqs.\ref{lehmann-a}
and in \ref{lehmann-b} can be expanded in two new basis $\{ \phi_l^+ \}$ and 
$\{ \phi_l^- \}$ depending on whether  the states are even or odd: 
\begin{eqnarray}
\hat{O}_r ~|\phi^+_l \ran &=& ~|\phi^+_l \ran ,\\
\hat{O}_r ~|\phi^-_l \ran &=& -|\phi^-_l \ran .
\label{transf-states}
\end{eqnarray}
The destruction operators for the  $\alpha$ and $\beta$ levels transform as
\begin{equation}
c_{\alpha \sigma} = \hat{O}^+_r ~c_{\beta \sigma} ~\hat{O}_r .
\label{transf-op}
\end{equation}
Suppose the ground state is an even state $|f^+ \ran$ (we confirmed this 
numerically). 
Then, there are two kinds of matrix elements, 
$\lan f^+| c_{\alpha \sigma} |\phi_l^+ \ran$ and 
$\lan f^+| c_{\alpha \sigma} |\phi_l^- \ran$. 
These two matrix elements must be compared with their corresponding 
matrix elements, calculated with the operator $c_{\beta \sigma}$.
Using Eqs.\ref{transf-states} and \ref{transf-op} we have
\begin{eqnarray}
\lan f^+| c_{\alpha \sigma} |\phi_l^+ \ran &=&
~\lan f^+| c_{\beta \sigma} |\phi_l^+ \ran , \\
\lan f^+| c_{\alpha \sigma} |\phi_l^- \ran &=&
-\lan f^+| c_{\beta \sigma} |\phi_l^- \ran .
\end{eqnarray}
Then, some of the poles of Eqs.\ref{lehmann-a} and \ref{lehmann-b}
carry weight with differents signs. When $\Delta V$ is different
from zero, the weights are different but the difference in the signs
remains.

\subsection{Interference between S=1/2 and S=1 Kondo states?}

Note that the conductance cancellations mainly occur for values of the gate
voltage where the charge of the dots is changing, for example
from 1 to 2. This suggests that an alternative way to visualize the
interference process is to imagine the ground state in this intermediate
regime as a linear combination of two Kondo states, one corresponding
to a spin 1/2 at the dot and the other to a spin 1 at the same dot. 
In fact, in the important range of gate voltages considered in our study, 
the charge at the dot is not sharply defined, thus this approximation
is reasonable. The interference
could arise from a destructive process between the transport
corresponding to each of these two nearly degenerate states.
This hypothesis will be analyzed more extensively in future studies.

\subsection{Localized States}

Note that the matrix element $\lan f^+| c^+_{r \sigma} |\phi_l^- \ran$
must be zero when $\Delta V$=$0$, since the operator 
$c^+_{r \sigma}$ does not change the symmetry of the state 
$|\phi_l^- \ran$ and the resulting state is orthogonal to 
$\lan f^+ |$.  
This fact produces {\it localized states} in the QD due to the
poles associated with the states $|\phi^-\ran$ (located at
$\omega = E_F-E_l$) which
 are present in $g_{\alpha \alpha}$ and $g_{\beta \beta}$
(the element $\lan f^+| c^+_{r \sigma} |\phi_l^- \ran$ does not 
appear in these Green functions), but they are $not$ present in the
Green function that transports electrons from one lead
to the other ($g_{\alpha r}$ or $g_{\l r}$).
This fact explains the results observed in the right frame of 
Fig. \ref{cluster-dif}. When $\Delta V$ is not zero, this is
no longer valid since $\hat{O}_r$ does not commute with $H_{\rm T}$.

\begin{figure}
\epsfxsize=8cm \centerline{\epsfbox{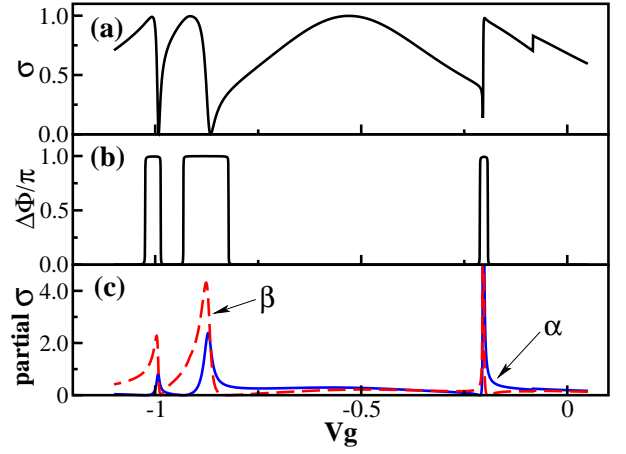}}
\caption{ Non-symmetric $\alpha$-$\beta$ case, where the Coulomb
repulsions in both levels are different i.e.
$U_\beta$=$U_\alpha + 0.1t$. Dips in the conductance are observed
even with $\Delta V$=$0$, due to the explicitly broken symmetry. More
details can be found in the text.}
\label{Ud-dif}
\end{figure}

One way to eliminate the symmetry between $\alpha$ and $\beta$ could
be by using a different diagonal Coulomb repulsion for both levels.
In Fig. \ref{Ud-dif} we can see the conductance, the phase difference, and
the partial conductances for the case with $\Delta V$=$0$ and
$U_\beta = U_\alpha + 0.1t$.
Note that the first discontinuity remains since for
one or zero electron in the QD, the different Coulomb repulsions
$U_\alpha$ and $U_\beta$ are not active and the symmetry is
not broken. In this case all the physics related to the 
poles associated with the symmetry $\alpha$-$\beta$ discussed
above could be applied and the discontinuities shown in 
Fig. \ref{fg1-1qd2l}(a) appears.

\section{ Conductance Cancellation in Similar Systems}
\label{sec:othercases}

In this section, we will discuss two other systems where a similar 
interference process is also operative.

%%%%%%%%%%%%%%%%%%%%%%%%%%%%%%%%%%%%%%%%%%%%%%%
%%%%%%%%  TWO QUANTUM DOTS IN T CONFIGURATION.
%%%%%%%%%%%%%%%%%%%%%%%%%%%%%%%%%%%%%%%%%%%%%%%
\subsection{ Two Coupled QD} \label{sec:VIa}
\begin{figure}
\vskip 0.5cm
\epsfxsize=8cm \centerline{\epsfbox{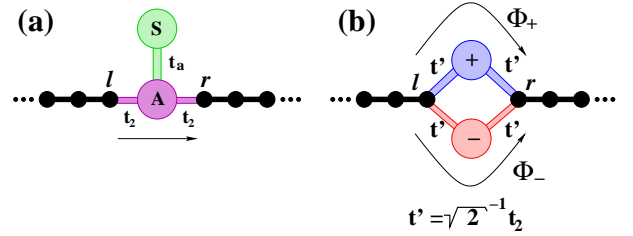}}
\vskip 0.25cm
\caption{ (a) Schematic representation of a system of two coupled QD,
one of them side-connected. In this case, each dot has only one level.
(b) Equivalent system after the
transformation Eqs.\ref{base+} and \ref{base-}. 
The two possible paths in the new basis are shown.}
\label{esquema}
\end{figure}

The first system to analyze consists of two coupled one-level 
quantum dots, one of them side-connected \cite{microelectronics,Hershfield1},
as shown schematically in Fig. \ref{esquema}(a).
The two dots ($A$ and $S$) are described by the Hamiltonian
$H$=$H_{\rm dots}$+$H_{\rm leads}$+$H_{\rm int}$, where
\begin{eqnarray}
H_{\rm dots} &=&  \sum_{\sigma,\lambda=\mbox{\tiny A,S}} \left\{
U/2 ~~n_{\lambda \sigma} ~n_{\lambda \bar{\sigma}}  ~+
 V_g ~~n_{\lambda \sigma} \right\} ~+~  \nonumber \\
 &&  t_a \sum_\sigma c_{\mbox{\tiny A}\sigma}^+
   ~c_{\mbox{\tiny S}\sigma} ~+~ \mbox{h.c.}, \nonumber \\
H_{\rm int} &=& t_2 \sum_{\sigma} ~c_{\mbox{\tiny A}\sigma}^+ 
\left[c_{l\sigma}~+~c_{r\sigma}\right] ~+~ \mbox{h.c.} \nonumber 
\end{eqnarray}
As these are two physically separated quantum dots, we will neglect 
the terms $U'$ and $J$ used in the previous section. 
Here, dot $A$ is the only one connected to 
the leads, by a matrix element $t_2$. $H_{\rm lead}$ is the same
as in previous sections.

Considering the two dots as a ``dimer'', $H_{\rm dots}$ can be rewritten
in the basis of bonding and anti-bonding states \cite{VApel},
\begin{eqnarray}  
c_{+\sigma} &=& \frac{1}{\sqrt{2}} \left[ c_{\mbox{\tiny A}\sigma} 
               ~+~ c_{\mbox{\tiny S}\sigma} \right] ,   \label{base+}\\
c_{-\sigma} &=& \frac{1}{\sqrt{2}} \left[ c_{\mbox{\tiny A}\sigma} 
                 ~-~ c_{\mbox{\tiny S}\sigma} \right] . \label{base-}                 
\end{eqnarray}
\noindent With this transformation, it is easy to visualize the two possible 
paths that produce the interference in the conductance 
(shown in Fig. \ref{esquema}). In the new basis, the matrix element
that connects the dimer state with the leads is $t'$=$1/\sqrt{2}~t_2$.
Note that the total charge in the two dots does not depend on the
basis used, since 
$\langle n_{\mbox{\tiny A}\sigma} + n_{\mbox{\tiny S}\sigma} \rangle$=$\langle n_{-\sigma} 
+ n_{+\sigma} \rangle$.
Also note that this bonding and anti-bonding states have a diagonal
single energy of value $V_{g-}$=$V_g - t_a$ and $V_{g+}$=$V_g + t_a$.
For this reason, the matrix element $t_a$ plays the same role that $2\Delta V$
had in the case of a quantum dot with two levels analyzed before.

\begin{figure}
\epsfxsize=8cm \centerline{\epsfbox{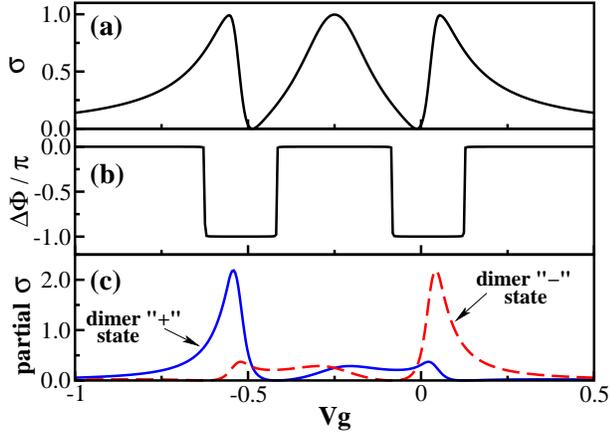}}
\caption{ Conductance (in units of $e^2/h$), phase difference and partial 
conductances as a function of the gate potential $V_g$, for the geometry 
shown in Fig.\ref{esquema}.
The parameters used, in units of $t$ (the bandwidth is $W=4t$), are 
$t_a=0.12$, $t_2=0.25$ and $U_i=0.5$.
In (a), two zeros in the conductance as a function of $V_g$ are shown.
(b) The phase difference $\Delta\Phi$ near these zeros jumps
to $-\pi$. In (c), the partials contributions
to the conductance $\sigma_-$ and $\sigma_+$ are given. 
For $V_g=-U_i/2$, we can observe a constructive interference process
where $\sigma_-=\sigma_+$ and $\Delta\Phi=0$.}
%One of the dips is shown in detail in fig \ref{detalle}.}
\label{fg1-2qd1l}
\end{figure}

As in the previous sections, we can define the conductance
as a function of the partial conductances through channels "$+$"
and ``$-$'',  as well as the phase difference between them as
$\sigma$=$\sigma_+ + \sigma_- + \sqrt{\sigma_+ \sigma_-} \cos(\Delta \Phi)$.
In Fig. \ref{fg1-2qd1l}, we show the results for the conductance,
phase and partial conductances $\sigma_+$ and $\sigma_-$ for the case
equivalent to that presented in Fig. \ref{fg1-1qd2l}b. 
Here, the dot coupling factor
$t_a$ used is $0.12t$. Again, a three-peaks structure in
the conductance can be observed. 
The zeros in the conductance appear when the phase
jumps to $-\pi$. This is a destructive interference process similar 
to that shown in Fig. \ref{DV0.2} for $\Delta V$=$0.2t$. 

\begin{figure}
\epsfxsize=8cm \centerline{\epsfbox{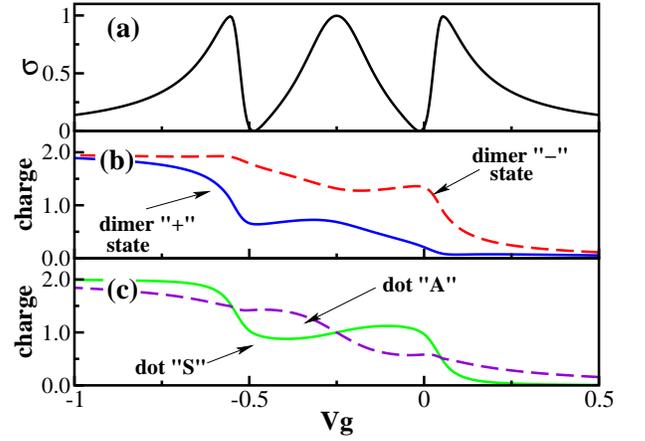}}
\caption{ Conductance (in units of $e^2/h$) and charge as a function 
of $V_g$, for
the same parameters used in Fig. \ref{fg1-2qd1l}. Shown are
(a) the conductance
vs. $V_g$; (b) the charge at the bonding and anti-bonding dimer 
states; (c) the charge at the real dots "A" and "S". 
}
\label{carga-2QD}
\end{figure}

As in section \ref{sec:one-dot}, the results presented here correspond
to the smallest cluster size $L=4$. In the final part of this
subsection the convergence with the size $L$ will be analyzed.
In Fig. \ref{carga-2QD}, the conductance and the charge
of the dimer states and dots are shown. The charge of the dimer states
presents the same behavior as the charge of the two-level dot studied 
before. The charge in the individual dots shows a slow increase for 
dot ``A'', and a  step-function-like behaviour  for dot ``S''. This indicates
that while the dot coupled to the leads is in a transition-valence 
regime and the charge enters slowly due to the Kondo resonance,
the other one is in a Coulomb-blockade regime.

\begin{figure}
\epsfxsize=8cm \centerline{\epsfbox{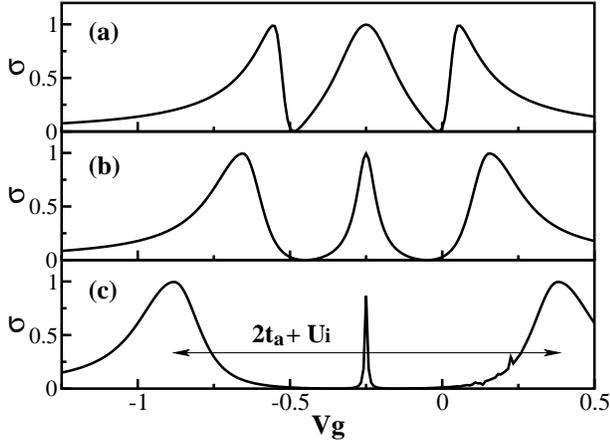}}
\caption{ Conductance (in units of $e^2/h$) for three different 
values of $t_a$. (a) $t_a$=$0.12$ (reproduced here for comparison), 
(b) $t_a$=$0.24$, and (c) $t_a$=$0.48$.
For $t_a$=$0$ the side coupled dot is not connected and the conductance
presents the characteristics of one QD (not shown). Increasing $t_a$,
the side coupled dot starts to participate in the transport,
and interference effects appear.
For large $t_a$, the system -- represented by 
the dimer states -- is equivalent
to one dot with two levels, separated by an energy $2t_a + U$.}
\label{cnd-ta}
\end{figure}

\begin{figure}
\epsfxsize=8cm \centerline{\epsfbox{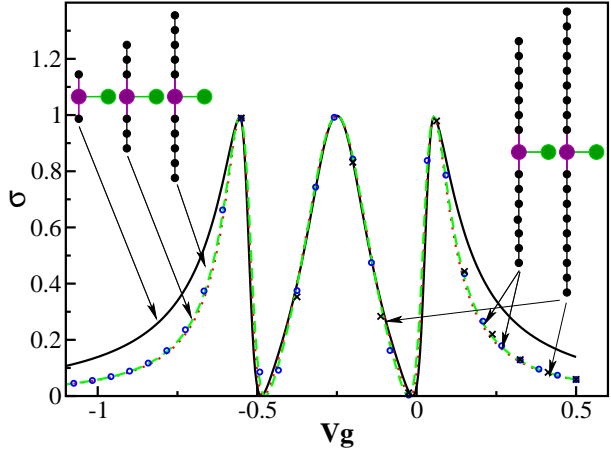}}
\caption{ Convergence of the conductance with the cluster 
size, for the case of the side-connected dot. The parameters are the 
same as in Fig. \ref{fg1-2qd1l}.}
\label{cconv}
\end{figure}

\begin{figure}
\epsfxsize=8cm  \centerline{\epsfbox{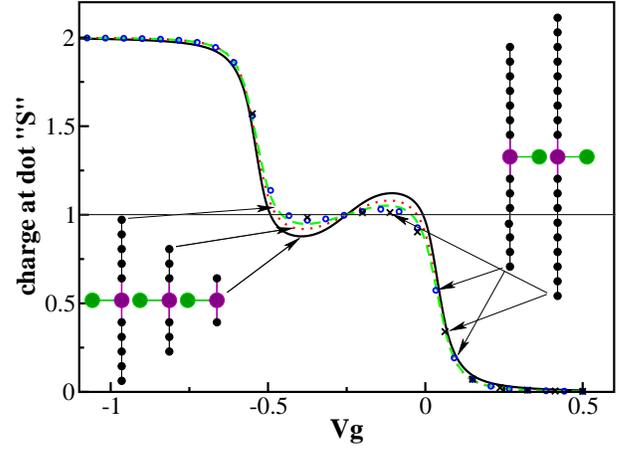}}
\caption{ Convergence study of the charge at dot ``S'' increasing the
size of the cluster exactly solved.
For the smaller cluster, we found important changes around $V_g$=$-U/2$,
but eventually the system converges to a plateaux for a sufficiently large
cluster. The parameters are the same as in Fig. \ref{fg1-2qd1l}.}
\label{cScrg}
\end{figure}

Increasing $t_a$ the dimer states are splitted and the two-dots
system can now be
 represented by one QD with two levels, with a large $\Delta V$$\sim$$2t_a$.
This is shown in Fig. \ref{cnd-ta}.
In the last two figures (Figs. \ref{cconv}, \ref{cScrg}) the
convergence with the cluster size is shown for the conductance and the charge.
As before, a fast convergence is observed.
In the case of the charge of dot ``S'', note that the wavy behavior
found for the smaller cluster is replaced by a ``plateaux''
when the cluster-size increased.

%%%%%%%%%%%%%%%%%%%%%%%%%%%%%%%%%%%%%%%%%%%%%%%
%%%%%%%%  TWO QUANTUM DOTS WITH TWO LEVELS.
%%%%%%%%%%%%%%%%%%%%%%%%%%%%%%%%%%%%%%%%%%%%%%%
\subsection{ Two coupled QD with two levels}

\begin{figure}
\vskip 0.25cm
\epsfxsize=7.5cm \centerline{\epsfbox{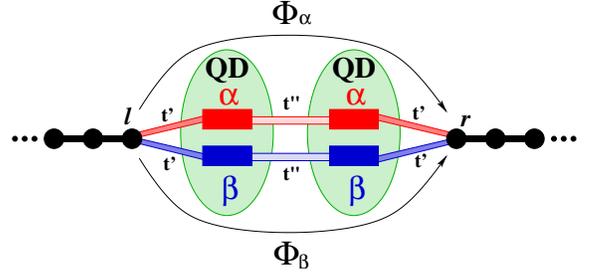}}
\vskip 0.25cm
\caption{ Schematic representation of the system of two 
coupled QD with two levels. The two possible paths 
carrying different phases $\Phi_\alpha$ and $\Phi_\beta$
are shown.}
\label{esquema-2QD2l}
\end{figure}

In this subsection, it will be analyzed
the case of two QD in series, each with two levels, as
schematically shown in Fig. \ref{esquema-2QD2l}. 
The Hamiltonian that describes this problem includes

\begin{eqnarray}
H_{\rm dots} &=& \sum_{i=1,2} \left\{ \sum_{\sigma,\lambda=\alpha \beta}  
U/2 ~n_{i\lambda \sigma} ~n_{i\lambda \bar{\sigma}}  ~+~ \right. \nonumber \\ &&
U'/2 ~\sum_{\sigma \sigma'}~n_{i\alpha \sigma} ~n_{i\beta \sigma'} ~+~  \nonumber \\
 && -J~ \sum_{\sigma \sigma'} ~c^+_{i\alpha \sigma} ~c_{i\alpha \sigma'} 
 ~c^+_{i\beta \sigma'}
 ~c_{i\beta \sigma} ~+~ \nonumber \\ &&
 \left. V_g ~\sum_{\sigma} \left[ ~n_{i\alpha \sigma} ~+~
   (V_g + \Delta V) ~n_{i\beta \sigma} \right] \right\} + \nonumber \\
 &&  \sum_{\sigma,\lambda=\alpha \beta} 
    t'' \left[ c^+_{1\lambda \sigma} c_{2\lambda \sigma} + \mbox{c.c.} \right] ,
\end{eqnarray}
\begin{equation}
H_{\rm int} = t' \sum_{\sigma, \lambda=\alpha,\beta} \left[ 
 c_{1\lambda \sigma}^+ c_{l 0\sigma} ~+~ c_{2\lambda \sigma}^+ c_{r 0\sigma} 
 ~+~ \mbox{c.c.} \right] ,
\end{equation}
where $c_{i\lambda\sigma}^+$ creates an electron at
the dot $i$ in the level $\lambda$ with spin $\sigma$. The Hamiltonian
which describes the leads is the same as before.
Then, the total Hamiltonian is
$H_{\rm T}$=$H_{\rm dot} + H_{\rm leads} + H_{\rm int}$.
As in the case of two QD with one level each, we can rewrite this
problem in the dimer-state basis. It is well known that in this kind 
of systems, a splitting in energy between the dimer states must occur 
\cite{busser,Wiel,AGeorges,2qd-exp}.
This separation is related to  the $t''$ matrix element.

\begin{figure}
\epsfxsize=8cm \centerline{\epsfbox{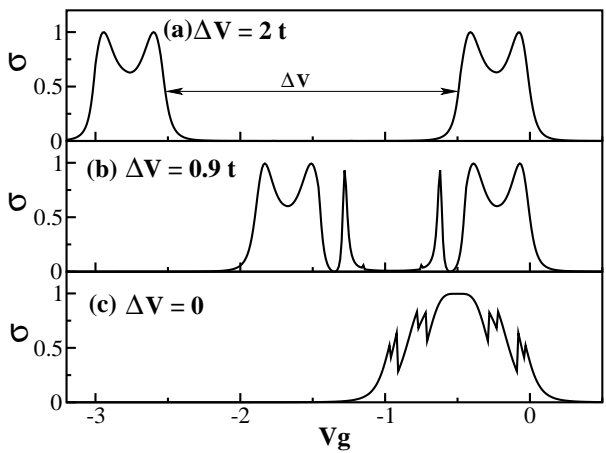}}
\caption{ Conductance of two coupled quantum dots, each with two levels,
 for three different cases of $\Delta V$
(large, intermediate, and zero). (a) When $\Delta V$ is the largest energy,
two structures separated by $\Delta V$ appear in the conductance,
similarly as for two coupled one-level QDs. (b) When $\Delta V$ decreases
these two structures start to overlap producing
dips in the conductance. (c) A more complex
conductance structure with several discontinuities is obtained
in the limit  $\Delta V$=$0 $.}
\label{fg1}
\end{figure}

As before we present the results corresponding to the smallest cluster
size. For this case $L=6$ that corresponds to two sites to describe
the ``leads'' and four levels to describe the two dots with two levels 
each.
Figure \ref{fg1} is the equivalent of Fig.\ref{fg1-1qd2l} for
the case of only one QD. The conductance for several $\Delta V$'s is
shown. As in the previous case, at large $\Delta V$ there are  two splitted
structures corresponding to the Kondo regime for the 2QD with one level.
An intermediate case $\Delta V$=$0.9t$, where these two structures
start to merge, is also shown.
When $\Delta V$=$0$, similar structures can be observed as found
 in previous cases,
where every discontinuity was splitted in two due to the dimer bonding
anti-bonding states. 

\begin{figure}
\epsfxsize=8cm \centerline{\epsfbox{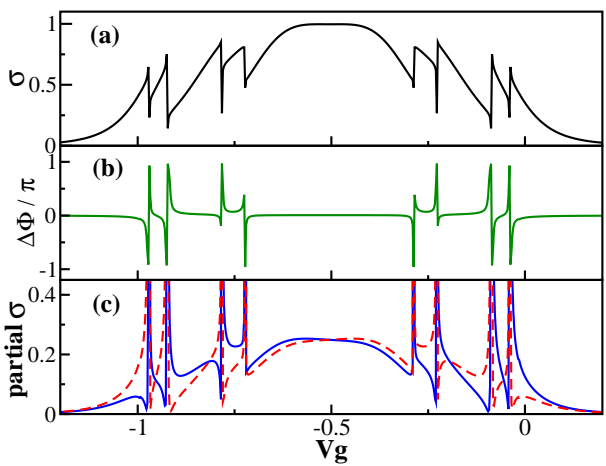}}
\caption{ Conductance (in units of $e^2/h$), phase, and partial conductances 
for a small $\Delta V$=$0.01 t$, in the case of two coupled two-levels QDs.
Typical dips in the conductance will be analyzed in the next figure.
Near the gate potential $V_g$=$-0.5$, a constructive interference occurs.
The phase is zero and the partial conductance becomes $\sim$0.25, inducing 
a constructive phase process for electron transport. }
\label{DV0.01c}
\end{figure}

\begin{figure}
\vskip 1cm
\epsfxsize=8cm \centerline{\epsfbox{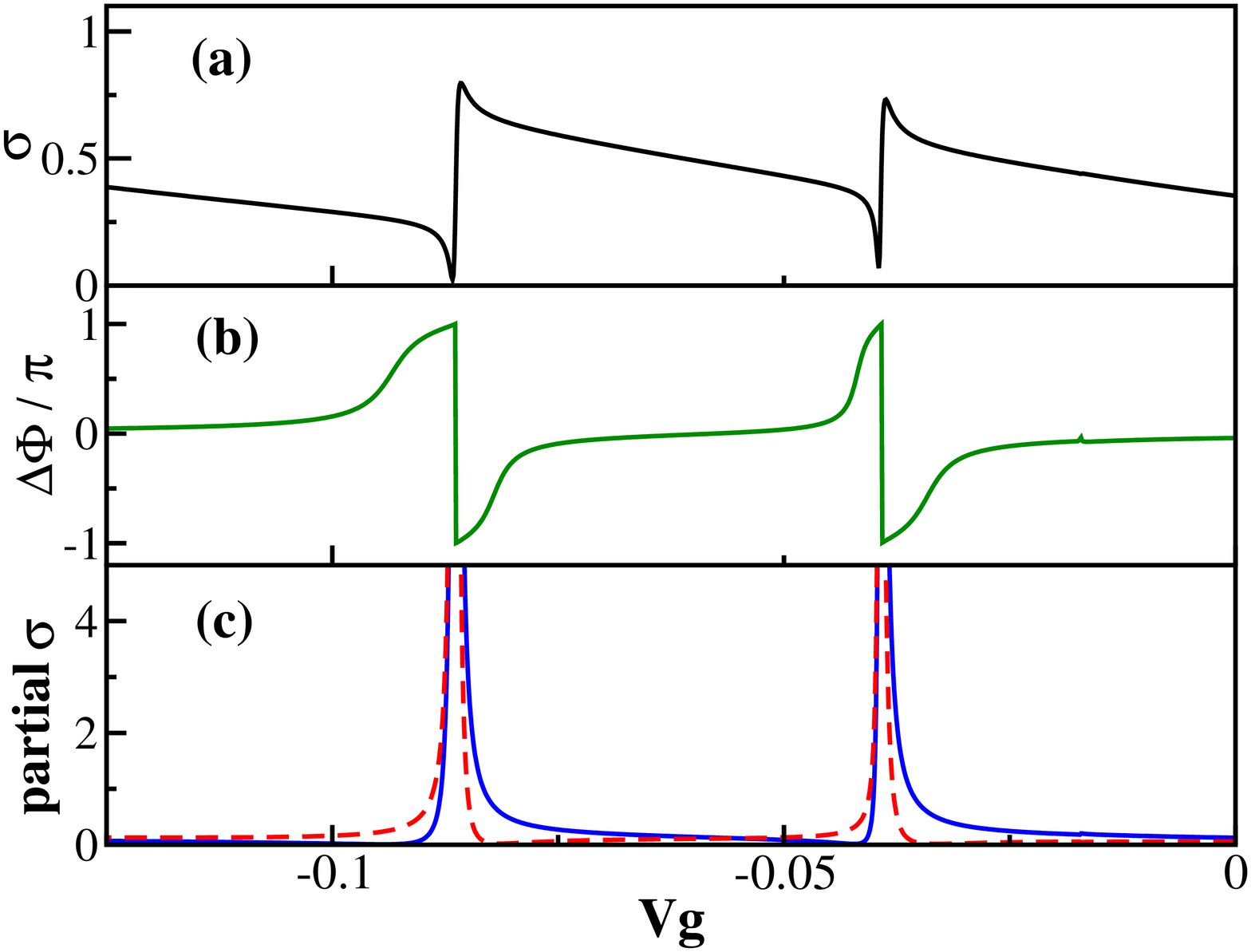}}
\caption{ First two dips in the conductance for the small $\Delta V$ case shown
in Fig. \ref{DV0.01c}. Both partial conductances ($\sigma_\alpha$ and 
$\sigma_\beta$) have a zero near the dips, but not at the same time. 
Then, the dip in the overall conductance is produced by a phase interference 
process induced by $\Delta \Phi$=$\pm \pi$.}
\label{DV0.01p}
\end{figure}

In Figures \ref{DV0.01c} and \ref{DV0.01p}, the conductance
and the phase for a very small $\Delta V$=$0.01t$ are presented. In this case
the  $\Delta V$=$0$ discontinuities are transformed into narrow
dips. In Fig. \ref{DV0.01p}, the first two dips are shown. They qualitatively correspond
to the splitting of the first dip in the 1QD case. 
The phase in Fig.\ref{DV0.01p} does not present the
abrupt jumps that were found for 1QD, but the dips nevertheless occur when the
phase takes the value $\Delta \Phi$=$\pm \pi$.

\section{ Conclusions}
\label{sec:conclusions}

In this paper, we have investigated systems where interference effects
in the conductance can arise in multilevel quantum dots. Although
there are no obvious different paths, as in a AB interferometer,
transport using different levels inside the QD can produce such interference.
Using an exact diagonalization plus embedding technique 
the phase difference between these paths was calculated.
When this phase difference is $\pm \pi$, a destructive
interference can appear. An important aspect of these results is that
the conductance cancellations can occur in the absence of
external magnetic fields -- usually used in AB experiments -- and
also without introducing at the Hamiltonian level any relative phase 
for the hopping amplitudes of the states involved. The
Coulombic interaction is the main responsible for the effect,
which can also be visualized as an interference between
$S_{\rm D}$=1 and 1/2 Kondo states.

Another important novel effect found in this investigation
 is the possibility of having
localized states inside the QDs. These states do not contribute
to the conductance but they modify the many-body
physics of the systems through Coulombic effects. 
These localized states induce discontinuities in
the conductance vs. gate voltage for particular couplings, while in
the most general case conductance cancellations are observed.

The results presented in this paper will hopefully provide motivation to
experimentalists to refine their measurements to search for conductance
cancellations. While our conductance vs. gate voltage curves should
be cautiously used to compare against experiments -- since the details of the 
results vary as a function of parameters, and in addition most results 
were obtained using only two active levels -- the ideas introduced here are 
expected to be robust. When multilevel quantum dots with interacting
electrons are considered, cancellations are the rule more than the exception.

Experimental realizations of quantum dots with many active
levels are possible. Recently, Aikawa {\it et al.} \cite{Aikawa} observed
that quantum-dot states can be classified into a small number of states
that are strongly coupled to the leads, and a large number of states weakly
coupled to those leads. The mixing between these states was found to be of
relevance to explain transport experiments. It is clear that many 
theoretical calculations must be revisited to incorporate the multiplicity
of levels in dots. This is also important
if transport properties through atoms are studied. This is 
more challenging
experimentally than the study of
relatively larger size quantum dots, but possible. In the case of
atoms, degeneracies are natural in open-shell atoms, and a variety of
interference processes as those observed here are possible.

\section{acknowledgments}

The authors acknowledge 
conversations with E.V. Anda, M.A. Davidovich, G. Chiappe and M.V. Apel.
Support was provided by the NSF grants DMR-0122523 and 0303348. Additional
funds have been provided by Martech (FSU).

%%%%%%%%%%%%%%%%%%%%%%%% References
%%%%%%%%%%%%%%%%%%%%%%%%%%%%%%%
%\begin{references}

%\end{multicols}


\begin{thebibliography}{99}
\bibitem{noteEOM} Note that the equation of motion for $G_{lr}$ take
 this simple form due the sites $l$ and $r$ have no many-body interactions.
\end{thebibliography}
\end{document}